\documentclass[epj,referee]{svjour}
\usepackage{amssymb}
\usepackage{graphicx}
\begin{document}
\title{Bundle vs. network conductivity of carbon nanotubes separated by type}

\author{H.M. T\'oh\'ati\inst{1} \and \'A. Pekker\inst{1}\thanks{\emph{Present address:} Center for Nanoscale Science and Engineering, Departments of Chemistry and Chemical \& Environmental Engineering, University of California, Riverside, CA 92521, U.S.A.}
\and B.\'A. Pataki\inst{1}
\and Zs. Szekr\'enyes\inst{1}
\and K. Kamar\'as\inst{1}\thanks{email: kamaras.katalin@wigner.mta.hu}
}% Do not remove
\institute{Institute for Solid State Physics and Optics, Wigner Research Centre for Physics, Hungarian Academy of Sciences, P.O. Box 49, H-1525 Budapest, Hungary}
\date{Received: date / Revised version: date}
\abstract{We report wide-range optical
investigations on transparent conducting networks made from
separated (semiconducting, metallic) and reference (mixed) single-walled carbon nanotubes, complemented by transport measurements. Comparing the intrinsic frequency-dependent conductivity of the nanotubes with that of the networks, we
conclude that higher intrinsic conductivity results in better transport properties, indicating that the properties of the nanotubes are at least as much important as the contacts. We find that HNO$_3$ doping offers a larger improvement in transparent conductive quality than separation. Spontaneous dedoping occurs in all samples but is most effective in films made of doped metallic tubes, where the sheet conductance returns close to its original value within 24 hours.
 % end of PACS codes
} %end of abstract
\maketitle
\section{Introduction}
\label{sec:intro}

One of the most promising applications of carbon-based new materials like carbon nanotubes or graphene is the area of transparent conducting layers \cite{gruner06,hecht11}. Carbon-based materials have many advantages over widely used oxides like indium tin oxide (ITO) in terms of better flexibility and no toxicity; however, their basic optical and electrical properties have not reached those of conventional transparent conductors so far. The field has been broadened recently by the possibility of separating nanotubes by electronic type \cite{green08,lu10}. By this method, not only highly enriched semiconducting or metallic networks can be prepared, but also extremely purified mixed samples \cite{green08}.

These samples offer a unique opportunity to study the role of intrinsic conductivity vs. intertube contacts in nanotube networks \cite{blackburn08,barnes08}. Intertube connections have been studied previously on junctions built from individual nanotubes \cite{fuhrer00} with similar or dissimilar electronic character (SS, MM or SM, respectively, where M stands for metallic and S for semiconducting tube). In macroscopic networks containing predominantly one type of nanotube, and by measuring both the transport and frequency-dependent
conductivity, these roles can be even more precisely determined. Impedance studies by Garrett et al. \cite{garrett10} showed that above a characteristic frequency of the order of kHz, the values reflect the intrabundle conductance instead of the combined values of bundles and junctions that is measured by the dc method. Thus infrared spectroscopy, which reaches down terahertz frequencies, clearly yields the intrinsic conductivity of the bundled nanotubes.

In this paper, we report the transparent and conducting properties of separated metallic and semiconducting single-walled nanotube (SWNT) films and compare them to those of an ultrahigh purity reference sample. The contactless measurement of the frequency-dependent conductivity and the four-point dc transport results on the same material can be directly related and the role of intrinsic conductivity vs. that of intertube contacts established. We also study the effect of doping on the transport and optical properties. In order to improve the conductivity and avoid chemical reactions at defects, we use mild p-doping by nitric acid vapor at room temperature \cite{kamaras10}. As this procedure is shown to result in a doped state unstable over time, it is only regarded as a proof-of-concept experiment. Nevertheless, two important practical questions can be addressed: 1. if we find a stable doping method, which kind of tubes should be used to obtain the best transparent conducting properties; 2. if stability is required over optimal conducting properties, which kind is the most stable against accidental doping? We find the answer by comparing optical transmittance and dc conductivity measurements on all three kinds of samples.

\section{Experimental}
\label{sec:exp}

We used very high purity SWNT samples commercialized by NanoIntegris \cite{nanoint}. Starting material was arc-discharge P2 by CarbonSolutions \cite{carbsolP2}. Separation of the nanotubes was performed by density gradient ultracentrifugation (DGU) \cite{green08}, resulting in separated metallic and semiconducting samples with 95\% nominal purity, and a mixed (reference) sample with 99\% SWNT content. For the latter, we assume a composition of 1/3 metallic and 2/3 semiconducting tubes. (The fact that all three samples underwent identical treatment starting from the original P2 material ensures that extrinsic factors due to sonication and other steps do not influence the comparison. These effects were extensively discussed in Ref.~\cite{blackburn08}.) The mean diameter of the nanotubes is 1.4 nm and their length varies from 100~nm to 4~$\mu$m. From the aqueous suspensions of surfactant-covered nanotubes we have prepared samples of different thickness using vacuum filtration \cite{wu04} through an acetone soluble filter. The thickness of the films was controlled by the applied amount of solution. These layers (three of each sample, differing in thickness) were subsequently relocated over a 1~cm x 1~cm x 1~mm quartz (suprasil) substrate. To remove any remnant of the solvent and traces of accidental atmospheric doping, the samples were annealed at 200$^{\circ}$C for 13h.

\begin{figure*}
\includegraphics{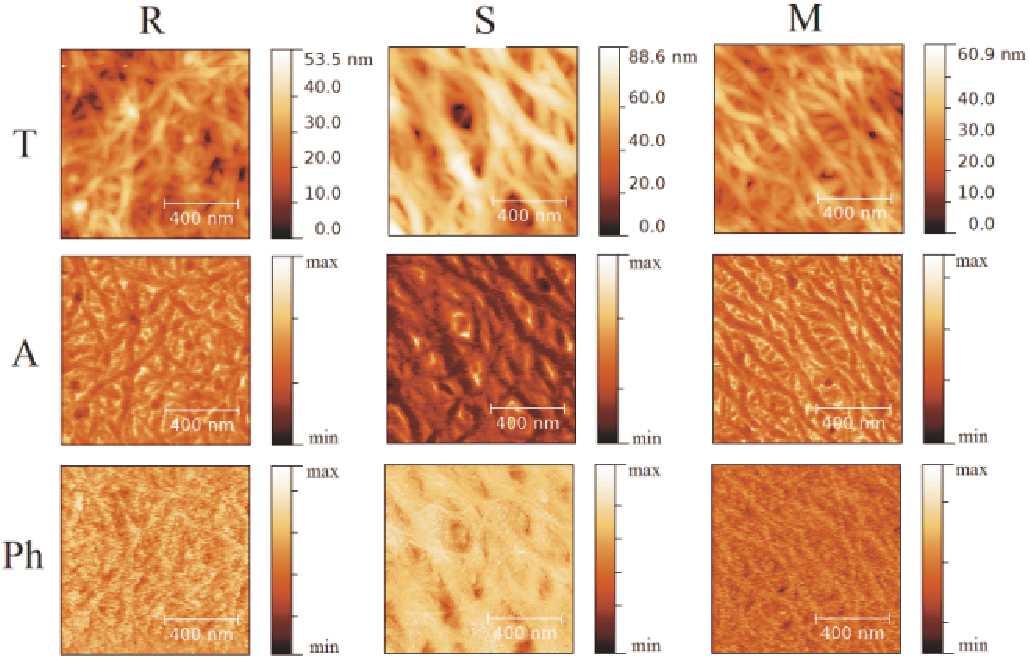}
\caption{Atomic force microscopy and near-field infrared amplitude (A) and phase (Ph) images of (a) reference sample (purified P2), (b) semiconducting sample and (c) metallic sample at 1000 cm$^{-1}$ laser frequency. The uniformity of the amplitude and phase images proves the sample purity.
\label{fig:AFM}}
\end{figure*}

Self-supporting thin films were prepared by stretching the nanotube layer over a hole created in a graphite disk. Using
this kind of samples enables us to measure transmission without the perturbation caused by substrates and to calculate easily the optical functions from transmission \cite{pekker11}.

Doping of the films was performed by subjecting them to nitric acid vapor overnight at room temperature. This mild treatment causes  hole doping in the $\pi$-electron system \cite{zhou05} without converting sp$^2$ carbon atoms into sp$^3$ by carboxylic group addition \cite{kamaras10}.

Scattering type near-field infrared microscopy (s-SNOM) data were taken with a NeaSNOM nano-FTIR instrument (Neaspec GmbH) using a quantum cascade  laser with 10.5 $\mu$m central wavelength. The s-SNOM technique is described in detail elsewhere \cite{keilmann09,ocelic06}. Briefly, the s-SNOM uses a metal coated AFM tip with the radius of curvature of ~20 nm enabling the operation at ultrahigh spatial resolution and near-field interaction between the illuminated tip and the sample \cite{amarie12,cvitkovic07}. Besides the usual AFM topography data the interferometric detection of the optical signal reveals local optical information including the absolute scattering efficiency (amplitude) as well as phase of the scattering \cite{keilmann09}. The observed optical contrasts are strongly related to the complex dielectric function of the studied material \cite{hillenbrand00}.

Wide range (far-infrared through ultraviolet) optical measurements were performed on the self-supporting nanotube networks using the following spectrometers: a Bruker IFS 66v/S Fourier-transform (FTIR) interferometer for the far-infrared
(FIR) and mid-infrared (MIR) range, a Bruker Tensor 37 FTIR in the near infrared (NIR), and an Ocean Optics QE65000 instrument for the ultraviolet-visible (UV-VIS) region. In the case of samples on quartz substrate, only data from the UV-VIS range were collected to obtain the transmission value at 550 nm (18180 cm$^{-1}$).

For each nanotube network on the quartz substrate, we measured the four-point resistivity using a Keithley 192 digital multimeter,  and calculated their sheet resistivity ($R_{\Box}$) applying the van der Pauw formula \cite{pauw58}. Small dots of colloidal silver were used for contacting the samples. All measurements were performed at room temperature.

Thickness of the self-supporting films was measured by atomic force microscopy following the procedure described in Ref. \cite{bekyarova05}. From the thickness and the spectra of each type of network, we determined the absorption coefficient at 550 nm and used these values to estimate the thickness of the samples used for resistivity, in order to determine the conductivity values. The absorption coefficients measured on the undoped samples are 3.39$\cdot$ 10$^4$ cm$^{-1}$ (R),
3.18$\cdot$ 10$^4$ cm$^{-1}$ (S), and 2.36$\cdot$ 10$^4$ cm$^{-1}$ (M), respectively. These values compare very well with those measured earlier on laser-deposited films prepared by the same procedure \cite{borondics06}.

\section{\label{sec:spectra}Frequency-dependent conductivity}

Figure \ref{fig:AFM} shows the local structure of the films of all three types. All films consist of bundles with about 50 nm average thickness as seen in both the topographic and optical images. The s-SNOM technique results in both amplitude and phase values of the scattered infrared light. The frequency range of the infrared laser is 955 - 1030 cm$^{-1}$, where the exciting radiation interacts with the free (Drude) carriers from the metallic nanotubes. Both amplitude and phase show a remarkable uniformity, proving the high purity of our samples. Additionally, the scattering amplitude from the metallic samples was found to be consistently higher than that from the semiconducting ones, at four distinct wavelengths, 970, 985, 1000 and 1015 cm$^{-1}$, indicating higher conductivity. A full account of s-SNOM measurements in a wider frequency range and using more sophisticated evaluation methods to obtain optical functions \cite{keilmann09} will be published elsewhere.

\begin{figure}[]
\includegraphics[width=8.5cm]{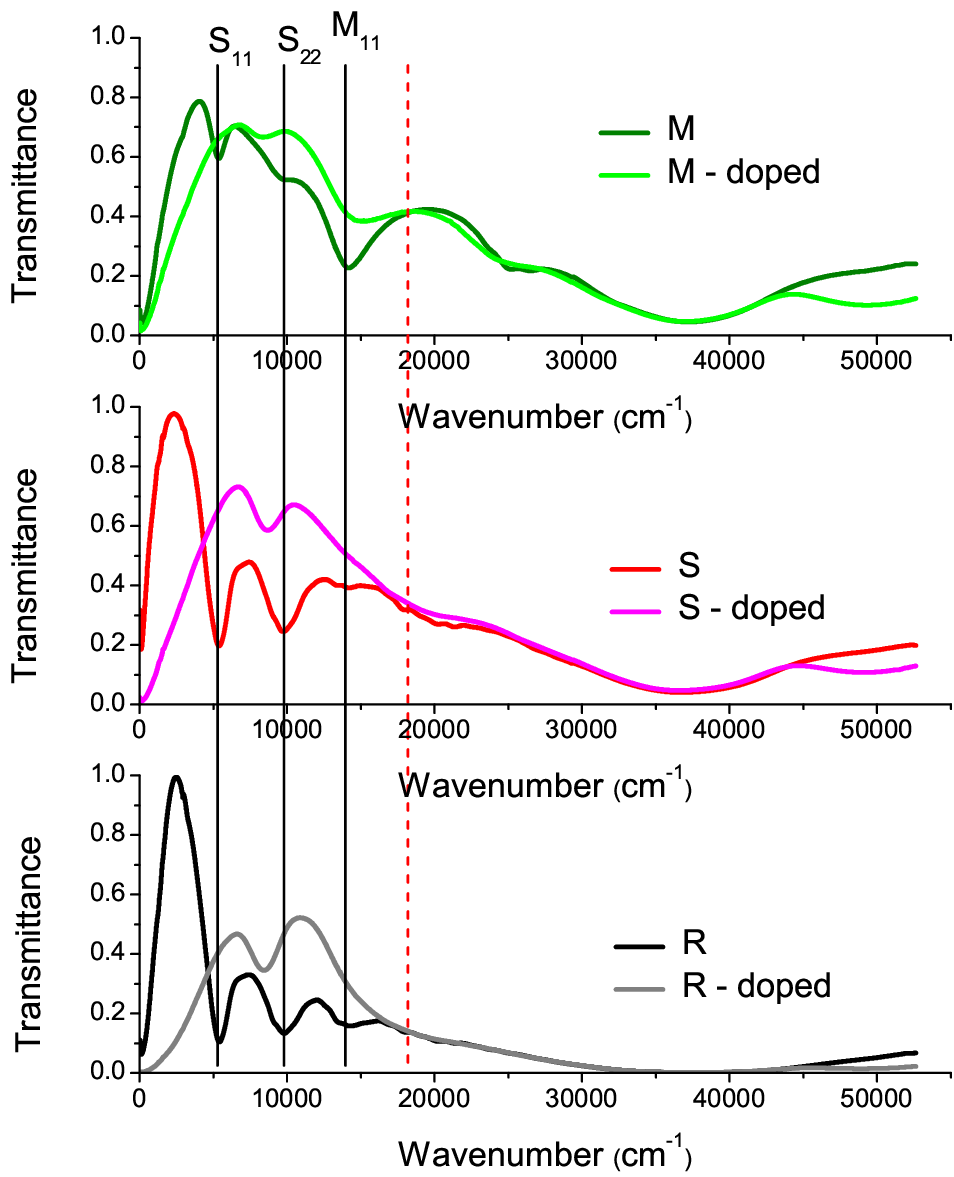}
\caption{\label{fig:trans} Frequency-dependent transmission spectra of self-supporting thin films in the FIR/UV range for metallic (M), semiconducting (S) and mixed reference (R) samples before (dark color) and after p-doping (light color). Solid black bars mark the first two semiconducting (S$_{11}$ and S$_{22}$) and the first metallic (M$_{11}$) transition; red dashed bar indicates the frequency corresponding to 550 nm wavelength.}

\end{figure}

Transmission spectra of self-supporting thin films of the mixed and separated nanotubes before and after doping are shown in
Fig. \ref{fig:trans}. These wide-range spectra prove that separation is effective: in the semiconducting sample the S$_{11}$ and S$_{22}$ transitions are very intense, while the M$_{11}$ transition can hardly be seen; in the case of the metallic sample the intensity of the M$_{11}$ transition is high and the peaks representing the transitions of semiconducting nanotubes are weak.

\begin{figure}[]
\includegraphics[width=8cm]{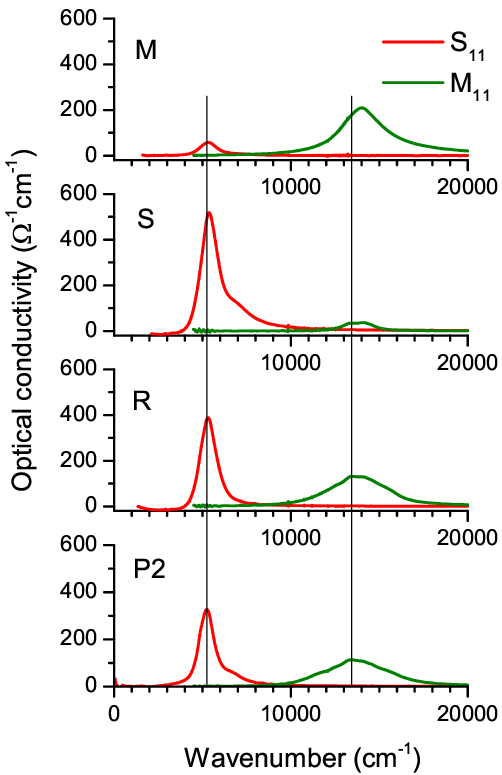}
\caption{\label{fig:sep}Optical conductivity in the S$_{11}$ and M$_{11}$ transition region for the three samples, indicating the relative semiconducting to metallic nanotube content. In the bottom panel, the same transitions for a commercial P2 sample are shown (Ref. \cite{pekker11}).}
\end{figure}

The intrinsic frequency-dependent conductivity can be calculated from the wide-range spectra and the thickness of the films by Kramers-Kronig transformation of the transmittance \cite{pekker11,borondics06} and contributions from individual transitions can be determined by the fitting procedure given in Ref. \cite{pekker11}. We show in Fig.~\ref{fig:sep} the optical conductivity curves corresponding to the S$_{11}$ and M$_{11}$ transitions,
respectively; these curves represent the envelopes of the transitions of semiconducting and metallic tubes of different diameter. For comparison, we include in Fig.~\ref{fig:sep} similar curves obtained for a commercial P2 sample \cite{pekker11}. The ratio of the areas A(S$_{11}$)/A(M$_{11}$), as expected, scales with the semiconductor/metal ratio: 1.03 and 0.98 for P2 and R,
respectively, 9.67 for S and 0.14 for M. Additional information included in these numbers is that upon DGU treatment, the composition of the reference sample does not change considerably. A slight blueshift of $\approx$ 100 cm$^{-1}$ occurs from P2 to R, and no further shift upon separation. The shift indicates that the purified mixed sample contains
slightly more small-diameter nanotubes than the original. The optical conductivity increases about 18 per cent upon the purification step for both the semiconducting and metallic transition.

Doping has the largest noticeable effect on the reference and semiconducting samples (Fig. \ref{fig:trans}), showing an increase of transmittance in the near infrared. The disappearance of the first and second interband transitions proves the high p-doping efficiency. (The appearance of the new transmission minimum between S$_{11}$ and S$_{22}$ is most probably originating in excitonic effects \cite{matsunaga11} and will be discussed elsewhere.) In addition, in all three samples the far-infrared transmission is decreasing, due to the free carriers introduced by doping \cite{borondics06}.

\begin{figure}[]
\includegraphics[width=8.5cm]{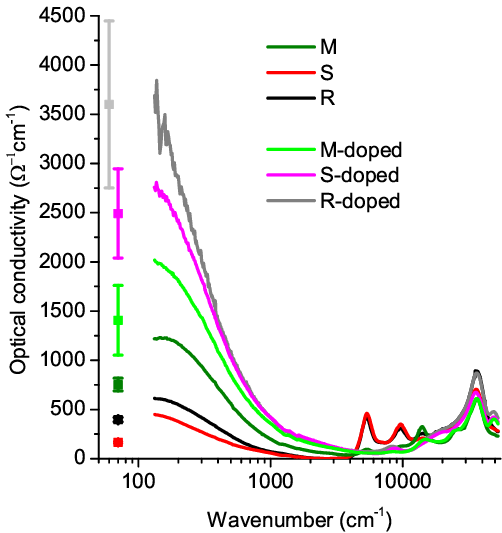}
\caption{\label{fig:cond} Frequency-dependent optical conductivity of doped and undoped nanotube films. The color code is the same as in Fig. \ref{fig:trans}. Note the logarithmic frequency scale. Squares with error bars represent measured dc conductivity values.}
\end{figure}

Figure \ref{fig:cond} shows the wide-range optical conductivity curves of all doped and undoped samples. The most striking difference (apart from the interband transitions) appears below 2000 cm$^{-1}$: all doped samples have a strong Drude contribution to the optical conductivity and therefore can be considered metals. For comparison we also show the dc conductivity of the samples from Fig. \ref{fig:cond1} obtained by averaging the respective data shown in Table \ref{tab:cond}. The low-frequency conductivity scales qualitatively with that obtained from the transport measurements, the latter being consistently lower, as expected for a heterogeneous structure involving contacts \cite{garrett10}. We regard this behavior as compelling evidence for the electronic structure of the nanotubes being mostly responsible for the conductivity enhancement upon doping, with contact effects playing a secondary role. A similar conclusion was drawn by Miyata  et
al. \cite{miyata08} who explained the selective conductivity enhancement by the differences in the electronic density of states of metallic and semiconducting tubes and their change upon doping.

Hole doping results in an increased carrier (hole) density and partially filled valence bands, leading to enhanced intrinsic conductivity of the nanotubes. The carrier density changes show up in the optical spectra \cite{kaza99,hennrich03}, with the free carrier absorption increasing and the interband transition intensities decreasing because of a decrease in the density of initial states for these transitions.

\begin{table}
\caption{Conductivity data for metallic, semiconducting and reference films of different thicknesses.}
\label{tab:cond}
\begin{tabular}{lccccc}
\hline\noalign{\smallskip}
& d &\multicolumn{2}{c}{Sheet conductance S} &\multicolumn{2}{c}{Conductivity $\sigma$ } \\
  & (nm) &\multicolumn{2}{c}{($10^{-3}\Box/\Omega$)} & \multicolumn{2}{c}{($\Omega^{-1}cm^{-1}$)} \\
 & & undoped & doped & undoped & doped \\
\noalign{\smallskip}\hline\noalign{\smallskip}
R-A & 50 & 2.04 & 14.3 & 405 & 2836 \\
R-B & 95 & 3.52 & 32.5 &  372 & 3435 \\
R-C & 177 & 7.33 & 80.0 & 415 & 4531 \\
S-A & 72 & 1.13 & 14.6 & 157 & 2028 \\
S-B & 135 & 2.38 & 33.9 & 176 & 2511 \\
S-C & 236 & 3.96 & 69.3 & 168 & 2936 \\
M-A &  92 & 6.4 & 9.94 & 698 & 1076 \\
M-B & 138 & 10.2 & 18.9 & 735 & 1362 \\
M-C & 374 & 30.9 & 66.7 & 827 & 1782 \\
\noalign{\smallskip}\hline
\end{tabular}
\end{table}

Conductivity data are summarized in Table \ref{tab:cond}. These data show that the highest relative increase (12.8 - 17.5) is observed for the semiconducting sample, followed by the reference (7 - 10.9) while the lowest increase is shown by the metallic tubes (1.5 - 2.2). (The thickness dependence of the conductivity enhancement is related to the percolation nature of conductance in the films \cite{bekyarova05,skakalova06}.) In absolute values, the doped reference sample gives the highest
sheet conductivity. Miyata et al. \cite{miyata08} performed sulfuric acid doping on laser ablated metallic and reference nanotube samples. They obtain a sheet conductance enhancement of 1.58 for their metallic and 19.8 for the reference sample. Comparing those with our numbers above and taking into account the difference in both starting material and doping agent, we find the agreement remarkable.

Studies on individual nanotube networks cited above \cite{fuhrer00} resulted in the observation that the resistivity of junctions between semiconducting and metallic nanotubes - which create Schottky barriers - is two orders of magnitude higher than the resistivity between tubes of the same electronic type. Among our six samples, the only one where such MS contacts are abundant, would be the undoped reference sample, since on doping all of them become metallic. Two observations should follow if intertube contacts played the key role in network conductivity: 1. in the undoped samples, both S and M type networks should have a higher dc conductivity than R, while at higher frequencies, the conductivity should scale with the metallic content; and 2. doping of the network R, where both metallic and semiconducting tubes occur, will result in dramatic decrease of intertube resistivity and consequent increase in dc conductivity. If intertube connections dominated during doping \cite{znidarsic13}, the effect for the reference sample would be much larger in the transport than in the optical data. Since our observations point to rather the opposite, we conclude that although intertube connections are important, they do not dominate transport properties of high-quality nanotube networks as these. Most probably, percolation channels exist even in the mixed networks between bundles that circumvent Schottky barriers, observed in  junctions between individual tubes.

A pure mechanical explanation for the conductivity enhancement upon acid treatment has also been suggested \cite{geng07}: according to this model, the only role the acidic dopant plays is to "clean" the intertube contacts from surfactant molecules present in the undoped sample. The presence of surfactant in the material has been, unfortunately, never proven, nor has any reaction been proposed between surfactant and acid. In our
wide-range spectra, the intense infrared absorption of these organic molecules should be observed if they were present in significant concentration. Also, our films are washed copiously with water after filtering and we believe that this treatment removes surfactants much more effectively and with much less residue than any unspecified reaction with an acid.

\section{\label{sec:networks} Practical consequences for applications}

\begin{figure}[]
\includegraphics[width=8cm]{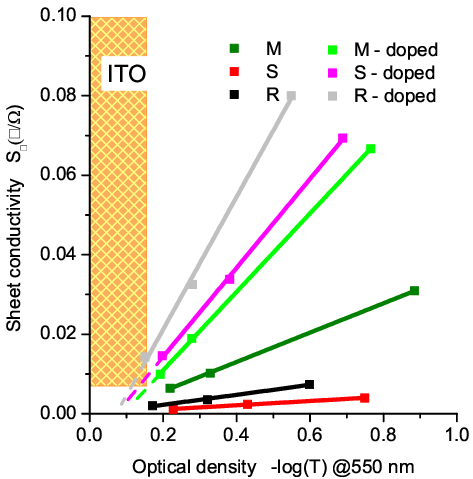}
\caption{\label{fig:cond1} dc sheet conductance S$_{\Box}$ vs optical density (-log T) at 550 nm for the metallic, semiconducting and mixed reference SWNT thin films after annealing and doping, respectively. The shaded area represents the application region of ITO.}
\end{figure}

For comparison of the transparent nanotube
networks, we applied a recently introduced figure of merit \cite{pekker10}, which is the inverse of that given by Jain and Kulshreshtha \cite{jain81}. This value is analogous to that defined by Gordon \cite{gordon00} but contains the transmission at a single wavelength instead of the integrated visible transmission, and similar to $\Phi_J$ in Ref.
\cite{Barnes12}. Fig. \ref{fig:cond1} shows the dc sheet
conductance S$_{\Box}$ as a function of optical density (\emph{-log T}) at 550 nm, the wavelength of choice for solar cell applications. Due to the fact that both the optical density and sheet conductivity are proportional to the
thickness of the film, for each sample the measured values
can be fitted with a linear function and the determination of
the thickness is not necessary. Higher slope of this
line indicates better quality of the film as a transparent conductor. The shaded area represents the region of ITO layers already used in technological applications (sheet resistance R$_{\Box} <$ 140 $\Omega/{\Box}$, $T >$ 0.7).\cite{green08} Extrapolating the values to lower thickness (dashed lines in Fig. \ref{fig:cond1}) indicates that both the doped reference
sample and the doped semiconducting sample reach the minimum parameters of the ITO region, while the doped metallic one barely misses it.

Figure \ref{fig:cond1} shows that doping causes the conductivity to increase in all samples, but the relative increase is different depending on electronic structure.
At the chosen wavelength of 550 nm, the increase in slope of the lines in Fig. \ref{fig:cond1}, therefore the increased performance as transparent conductor, is determined by the change in conductivity rather than absorbance. From the optical density values in Fig. \ref{fig:cond1}, a change within 15 per cent in absorbance can be deduced. In the near infrared, however, the transmittivity of the samples with substantial semiconductor content rises dramatically (Fig. \ref{fig:trans}). This means that for near-infrared applications, carbon nanotubes would be even more hopeful than for visible ones.

\begin{figure}[]
\includegraphics[width=8cm]{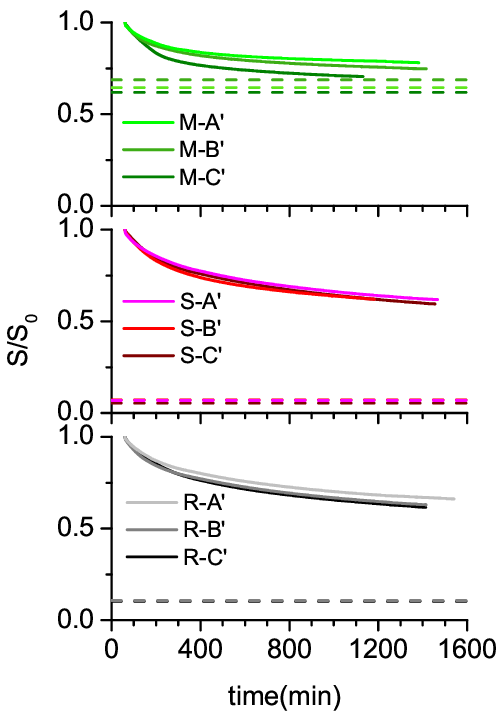}
\caption{\label{fig:tdep} Time dependence of the sheet conductivity of metallic, semiconducting and reference samples after doping with nitric acid. Data have been scaled to the sheet conductivity value at 60 minutes after removal from nitric acid vapor. The dashed lines indicate the sheet
conductivity of the undoped films. The samples numbered A', B' and C', respectively, increase in thickness in that order.}
\end{figure}

The most common practical problem with gas-phase doping is its reversibility, i.e. the dedoping process which usually starts as soon as the doping agent is removed. Indirectly, this effect is seen in Fig. \ref{fig:cond} where the error bars in the dc conductivity of the doped samples are significantly larger than those for the undoped samples. We have followed the dedoping process by measuring the sheet conductance for 24 hours after
removing the samples from the nitric acid vapor. Figure \ref{fig:tdep} shows these curves for a set of nine samples similar to those in Fig. \ref{fig:cond1}. The values before doping (relative to the as-doped state) are shown as dashed lines on each plot. All three types of samples show a behavior tending towards saturation which can be described by the sum of two exponential functions of time, but the parameters do not
seem to have real physical significance and therefore we do not discuss them here. The most striking feature of Fig. \ref{fig:tdep} is that contrary to the semiconducting and mixed samples, which approach a much higher saturation conductivity than before doping, the metallic samples seem to return to their original conductance in about a day. Doping is thus not
very effective for these materials, but on the other hand, they seem to be the most stable against accidental doping.

\section{\label{sec:concl}Conclusion}

Conclusions from our experiments can be drawn on two levels. One is the role of bundles vs. contacts in network conductivity.  Our frequency-dependent conductivity data follow the same trend as the values deduced from transport, meaning that the overall network conductivity is primarily influenced by what is happening in the bundles. By acid doping, the intrinsic conductivity increases considerably, especially for semiconducting nanotubes; this effect is followed by a similar increase in sheet conductivity. No anomaly is detected in either the intrinsic bundle conductivity or the overall network conductivity when going from the mixed (undoped R) to the all-metallic (doped R) network, which would be expected if a qualitative change in the contacts from Schottky to tunnel junctions \cite{fuhrer00} would determine the macroscopic properties. These facts prove that
although contacts play a crucial part in increasing conductivity of the networks, they are not the exclusive reason for improved electrical properties; in a macroscopic sample, inter-bundle pathways containing mostly tunnel-type junctions can be responsible for the conducting mechanism. 

The other type of conclusion concerns the application possibilities of separated and doped nanotube networks, respectively. If one chooses charge doping to increase the dc conductivity of carbon nanotubes, separation by type does not improve the results. Doped non-separated nanotubes can be best applied to substitute ITO for technological applications using visible light. However, undoped nanotubes of metallic type show other advantages: they have better conductance properties than either semiconducting or mixed ones, moreover, when doped, they recover their initial conductivity within less than 24 hours and are therefore much more stable against incidental doping, as already stated in Ref. \cite{miyata08}.

\begin{acknowledgement}
This work was supported by a joint project of the Hungarian Scientific Research Fund (OTKA) and the Austrian Science Fund (FWF) under Grant No. ANN 107580.
\end{acknowledgement}

%
% BibTeX users please use
\bibliographystyle{epj}
\bibliography{Tohati_Bundles}

\begin{thebibliography}{33}

\bibitem{gruner06}
G.~Gruner, J. Mater. Chem. \textbf{16}, 3533 (2006)

\bibitem{hecht11}
D.S. Hecht, L.~Hu, G.~Irvin, Adv. Mater. \textbf{23}, 1482 (2011)

\bibitem{green08}
A.A. Green, M.C. Hersam, Nano Lett. \textbf{8}, 1417 (2008)

\bibitem{lu10}
F.~Lu, M.J. Meziani, L.~Cao, Y.P. Sun, Langmuir \textbf{27}, 4339 (2011)

\bibitem{blackburn08}
J.L. Blackburn, T.M. Barnes, M.C. Beard, Y.H. Kim, R.C. Tenent, T.J. McDonald,
  B.~To, J.~Coutts, M.J. Heben, ACS Nano \textbf{2}, 1266 (2008)

\bibitem{barnes08}
T.M. Barnes, J.L. Blackburn, J.~van~de Lagemaat, T.J. Coutts, M.J. Heben, ACS
  Nano \textbf{2}, 1968 (2008)

\bibitem{fuhrer00}
M.S. Fuhrer, J.~Nygard, L.~Shih, M.~Forero, Y.G. Yoon, M.S.C. Mazzoni, H.J.
  Choi, J.~Ihm, S.G. Louie, A.~Zettl et~al., Science \textbf{288}, 494 (2000)

\bibitem{garrett10}
M.P. Garrett, I.N. Ivanov, R.A. Gerhardt, A.A. Puretzky, D.B. Geohegan, Appl.
  Phys. Lett. \textbf{97}, 163105 (2010)

\bibitem{kamaras10}
K.~Kamar\'as, {\'A}.~Pekker, B.~Botka, H.~Hu, S.~Niyogi, M.E. Itkis, R.C.
  Haddon, Phys. Stat. Sol. (b) \textbf{247}, 2754 (2010)

\bibitem{nanoint}
~www.nanointegris.com

\bibitem{carbsolP2}
~www.carbonsolution.com/products/p2-swnt.html

\bibitem{wu04}
Z.~Wu, Z.~Chen, X.~Du, J.M. Logan, J.~Sippel, M.~Nikolou, K.~Kamaras, J.R.
  Reynolds, D.B. Tanner, A.F. Hebard et~al., Science \textbf{305}, 1273 (2004)

\bibitem{pekker11}
{\'A}.~Pekker, K.~Kamar\'as, Phys. Rev. B \textbf{84}, 075475 (2011)

\bibitem{zhou05}
W.~Zhou, J.~Vavro, N.M. Nemes, J.E. Fischer, F.~Borondics, K.~Kamar\'as, D.B.
  Tanner, Phys.\ Rev.\ B \textbf{71}, 205423 (2005)

\bibitem{keilmann09}
F.~Keilmann, R.~Hillenbrand, in \emph{Nano-Optics and near-field optical
  microscopy}, edited by D.~Richards, A.~Zayats (Artech House: Boston, London,
  2009), p. 235

\bibitem{ocelic06}
N.~Ocelic, A.~Huber, R.~Hillenbrand, Appl. Phys. Lett \textbf{89}, 101124
  (2006)

\bibitem{amarie12}
S.~Amarie, P.~Zaslansky, Y.~Kajihara, E.~Griesshaber, W.~Schmahl, F.~Keilmann,
  Beilstein J. Nanotechnol. \textbf{3}, 312 (2012)

\bibitem{cvitkovic07}
A.~Cvitkovic, N.~Ocelic, R.~Hillenbrand, Nano Lett \textbf{7}, 3177 (2007)

\bibitem{hillenbrand00}
R.~Hillenbrand, F.~Keilmann, Phys. Rev. Lett \textbf{85}, 3029 (2000)

\bibitem{pauw58}
J.~van~der Pauw, Philips Res. Rep. \textbf{13}, 1 (1958)

\bibitem{bekyarova05}
E.~Bekyarova, M.E. Itkis, N.~Cabrera, B.~Zhao, A.~Yu, J.~Gao, R.C. Haddon, J.
  Am. Chem. Soc. \textbf{127}, 5990 (????)

\bibitem{borondics06}
F.~Borondics, K.~Kamar\'as, M.~Nikolou, D.B. Tanner, Z.~Chen, A.G. Rinzler,
  Phys. Rev. B \textbf{74}, 045431 (2006)

\bibitem{matsunaga11}
R.~Matsunaga, K.~Matsuda, Y.~Kanemitsu, Phys.\ Rev.\ Lett. \textbf{106}, 037404
  (2011)

\bibitem{miyata08}
Y.~Miyata, K.~Yanagi, Y.~Maniwa, H.~Kataura, J. Phys. Chem. C \textbf{112},
  3591 (2008)

\bibitem{kaza99}
S.~Kazaoui, N.~Minami, R.~Jacquemin, H.~Kataura, Y.~Achiba, Phys.\ Rev.\ B
  \textbf{60}, 13339 (1999)

\bibitem{hennrich03}
F.~Hennrich, R.~Wellmann, S.~Malik, S.~Lebedkin, M.M. Kappes, Phys.\ Chem.\
  Chem.\ Phys. \textbf{5}, 178 (2003)

\bibitem{skakalova06}
V.~Sk\'akalov\'a, A.B. Kaiser, Y.S. Woo, S.~Roth, Phys.\ Rev.\ B \textbf{74},
  085403 (2006)

\bibitem{znidarsic13}
A.~Znidarsic, A.~Kaskela, P.~Laiho, M.~Gaberscek, Y.~Ohno, A.G. Nasibulin,
  E.~I.Kauppinen, A.~Hassanien, J. Phys. Chem. C \textbf{117}, 13324 (2013)

\bibitem{geng07}
H.Z. Geng, K.K. Kim, K.P. So, Y.S. Lee, Y.~Chang, Y.H. Lee, J. Am. Chem. Soc.
  \textbf{129}, 7758 (2007)

\bibitem{pekker10}
{\'A}.~Pekker, K.~Kamar\'as, J. Appl. Phys. \textbf{108}, 054318 (2010)

\bibitem{jain81}
V.K. Jain, A.P. Kulshreshtha, Sol. Energ. Mater. \textbf{4}, 151 (1981)

\bibitem{gordon00}
R.G. Gordon, MRS Bull. \textbf{25}, 52 (2000)

\bibitem{Barnes12}
T.M. Barnes, M.O. Reese, J.D. Bergeson, B.A. Larsen, J.L. Blackburn, M.C.
  Beard, J.~Bult, J.~van~de Lagemaat, Adv. Energ. Mater. \textbf{2}, 353 (2012)

\end{thebibliography}

\end{document}